\documentclass[a4paper,twocolumn,11pt
,accepted=2026-03-31]{quantumarticle}
\pdfoutput=1
\usepackage[dvipsnames]{xcolor}

\usepackage[utf8]{inputenc}
\usepackage[english]{babel}
\usepackage[T1]{fontenc}
\usepackage{amssymb}
\usepackage{amsmath}
\usepackage{braket}
\usepackage{hyperref}

\usepackage{xspace}
\usepackage{tikz}
\usepackage{lipsum}

\newcommand{\RM}{\texttt{RandomMeas.jl}\xspace}

\newcommand{\change}[1]{{#1}}

\usepackage{listings}
\definecolor{codegreen}{rgb}{0,0.6,0}
\definecolor{codegray}{rgb}{0.5,0.5,0.5}
\definecolor{codepurple}{rgb}{0.58,0,0.82}
\definecolor{codered}{rgb}{0.796,0.235,0.2}
\definecolor{backcolour}{rgb}{0.95,0.95,0.92}
\definecolor{dodgerblue4}{rgb}{0.06, 0.31, 0.55}

\lstdefinelanguage{Julia}{
    keywords=[1]{abstract, break, catch, const, continue, do, else, elseif, end, export, false, for, function, if, import, let, local, macro, module, quote, return, struct, true, try, using, while, where},
    keywords=[2]{Bool, Char, Dict, Float64, Int, String, Array, AbstractArray, Vector, Set, Tuple, NamedTuple, QuantumObject, QuantumObjectEvolution},
    keywords=[3]{dummy_start, versioninfo, Qobj, QobjEvo, tensor, basis, fock, destroy, qeye, sigmax, sigmay, sigmaz, sigmam, sigmap, range, sesolve, mesolve, mcsolve, ssesolve, smesolve, sqrt, sin, cos, steadystate, steadystate_fourier, dfd_mesolve, dsf_mesolve, coherent, sum, size, dropdims, H_dsf, c_ops_dsf, e_ops_dsf, cu, CuVector, wigner, normalize, addprocs, SlurmManager, set_num_threads, Lattice, mapreduce, multisite_operator, DissipativeIsing, Val, ntuple, EnsembleSplitThreads, rmprocs, workers, gradient, my_f_mesolve, real, expect, BacksolveAdjoint, EnzymeVJP, liouvillian, dummy_end},
    sensitive=true,  
    morecomment=[l]\#,%
    morecomment=[n]{\#=}{=\#},%
    morestring=[s]{"}{"},%
    literate={
        {\\}{{{\color{codered}\lstum@backslash}}}{1} {\{}{{{\color{codered}\{}}}{1}
        {\}}{{{\color{codered}\}}}}{1} 
        {\%}{{{\color{codered}\%}}}{1} {&}{{{\color{codered}\&}}}{1}
        {+}{{{\color{codered}+}}}{1}
        {*}{{{\color{codered}*}}}{1}
        {/}{{{\color{codered}/}}}{1}
        {'}{{{\color{codered}\textquotesingle}}}{1}
        {=}{{{\color{codered}=}}}{1}
        {<}{{{\color{codered}<}}}{1} 
        {=>}{{{\color{codered}=>}}}{1} {|>}{{{\color{codered}|>}}}{1}
        {==}{{{\color{codered}==}}}{1}
        {>}{{{\color{codered}>}}}{1} {?}{{{\color{codered}?}}}{1}
        {julia>}{{{\color{codered}julia>}}}{1}
        {^}{{{\color{codered}\textasciicircum}}}{1} {|}{{{\color{codered}|}}}{1}
        {~}{{{\color{codered}\textasciitilde{}}}}{1}
        {×}{{{\color{codered}$\times$}}}{1} {⋅}{{{\color{codered}$\cdot$}}}{1}
        {π}{{$\pi$}}1
        {ω}{{$\omega$}}1
        {α}{{$\alpha$}}1
        {β}{{$\beta$}}1
        {σ}{{$\sigma$}}1
        {ψ}{{$\psi$}}1
        {κ}{{$\kappa$}}1
        {γ}{{$\gamma$}}1
        {Δ}{{$\Delta$}}1
        {φ}{{$\phi$}}1
        {ξ}{{$\xi$}}1
    }
}

\lstdefinestyle{mainstyle}{
    language=Julia,
    backgroundcolor=\color{backcolour},   
    commentstyle=\color{gray},    
    keywordstyle=\color{blue}\bfseries,  
    keywordstyle=[2]\color{teal},  
    keywordstyle=[3]\color{dodgerblue4},  
    numberstyle=\tiny\color{codegray},
    stringstyle=\color{codepurple},
    basicstyle=\ttfamily\footnotesize,
    columns=fullflexible,
    upquote=true,
    breakatwhitespace=false,         
    breaklines=true,                 
    captionpos=b,                    
    keepspaces=true,                 
    numbers=left,                    
    numbersep=5pt,                  
    showspaces=false,                
    showstringspaces=false,
    showtabs=false,                  
    tabsize=2,
}

\lstdefinestyle{tablestyle}{
    language=Julia,
    backgroundcolor=\color{backcolour},
    keywordstyle=[2]\color{black},  
    keywordstyle=[3]\color{black},  
    basicstyle=\ttfamily\footnotesize,
    columns=fullflexible,
    upquote=true,
    breakatwhitespace=false,         
    breaklines=true,                 
    captionpos=b,                    
    keepspaces=true,                 
    numbers=left,                    
    numbersep=5pt,                  
    showspaces=false,                
    showstringspaces=false,
    showtabs=false,                  
    tabsize=2,
}

\begin{document}

\title{RandomMeas.jl: A Julia Package for Randomized Measurements in Quantum Devices}
\author{Andreas Elben}
\affiliation{PSI Center for Scientific Computing, Theory and Data, Paul Scherrer Institute, 5232 Villigen PSI, Switzerland}
\affiliation{ETH Zürich - PSI Quantum Computing Hub, Paul Scherrer Institute, 5232 Villigen PSI, Switzerland}
\orcid{https://orcid.org/0000-0003-1444-6356}
\author{Beno\^it Vermersch}
\affiliation{Univ.  Grenoble Alpes, CNRS, LPMMC, 38000 Grenoble, France}
\affiliation{Quobly, 38000 Grenoble, France}
\orcid{https://orcid.org/0000-0001-6781-2079}
\maketitle

\begin{abstract}
We introduce \texttt{RandomMeas.jl}, a modular and high-performance open-source software package written in Julia for implementing and analyzing randomized measurement protocols in quantum computing. Randomized measurements provide a powerful framework for extracting properties of quantum states and processes such as expectation values, entanglement, and fidelities using simple experimental procedures combined with classical post-processing, most prominently via the classical shadow formalism. \texttt{RandomMeas.jl} covers the full randomized measurement workflow, from the generation of measurement settings for use on a quantum computer, the optional classical simulation of randomized measurements with tensor networks, to a suite of estimators for physical properties based on classical shadows. The package includes advanced features such as robust and shallow shadow techniques, batch estimators, and built-in statistical uncertainty estimation. Its unified, composable design enables the scalable application and further development of randomized measurements protocols across theoretical and experimental contexts.
\end{abstract}

\section{Introduction}

Randomized measurement (RM) protocols have emerged as a widely used experimental technique for extracting expectation values, entanglement properties, fidelities, and even full process information from quantum systems~\cite{elben_randomized_2022}. They offer a unifying strategy for characterizing many-qubit states using modest hardware resources \cite{huang_predicting_2020} and have been demonstrated across various platforms including trapped ions \cite{brydges_probing_2019,joshi2020quantum,Zhu_2022,joshi_observing_2024}, photonic devices \cite{zhang2021experimental,Struchalin_2021}, and superconducting circuits \cite{satzinger_realizing_2021,vitale_robust_2024,hu_demonstration_2025,Dong_2025,andersen_thermalization_2025,votto_learning_2026}.

An RM experiment consists of sampling a set of random unitaries \(\{U_\alpha\}_{\alpha=1}^{N_{\mathrm U}}\) from a prescribed ensemble, applying each unitary to the quantum state of interest, and performing a projective measurement in the computational basis to obtain a bitstring \(\mathbf{s} \in \{0,1\}^{N}\) \cite{vanenk2012measuring,ohliger2013efficient,elben_renyi_2018,elben_statistical_2019}. Repeating this process \(N_{\mathrm M}\) times per unitary results in a dataset of \(N_{\mathrm U} \times N_{\mathrm M}\) bitstrings. These data are then analyzed classically to extract the physical properties of interest, often using the classical shadow formalism \cite{huang_predicting_2020}. Crucially, the measurement protocol is independent of the final observable, enabling the guiding principle of RM: \emph{“measure first, ask questions later”} \cite{elben_randomized_2022}.

While the data acquisition procedure is deliberately simple, efficiently processing the resulting large datasets—particularly for intermediate- and large-scale quantum systems—requires dedicated classical software. In parallel, recent years have witnessed a surge of theoretical advances in classical shadows and RM techniques, greatly enhancing the capabilities of RM-based methods. However, their classical software implementations are often developed independently, resulting in incompatible interfaces and code structures.

\change{\texttt{RandomMeas.jl} addresses these challenges by offering a unified, modular, and high-performance \textsc{Julia} package that supports the entire RM workflow, and takes advantage of the ITensors library~\texttt{ITensors.jl}
 for performing tensor-network operations.}
The package includes sampling measurement settings, optionally simulating the measurement process via tensor networks (with decoherence), constructing various types of classical shadows, applying robust error mitigation routines, and computing a range of statistical estimators with built-in uncertainty quantification. The package emphasizes standardization and modularity, making RM methods—including recent theoretical innovations—more accessible, reusable, and extensible for the broader quantum research community. The current release represents a first step, with further features planned in future updates, as outlined in  the outlook section~\ref{sec:outlook}.

\begin{figure*}
\begin{center}
\includegraphics[width=0.8\textwidth]{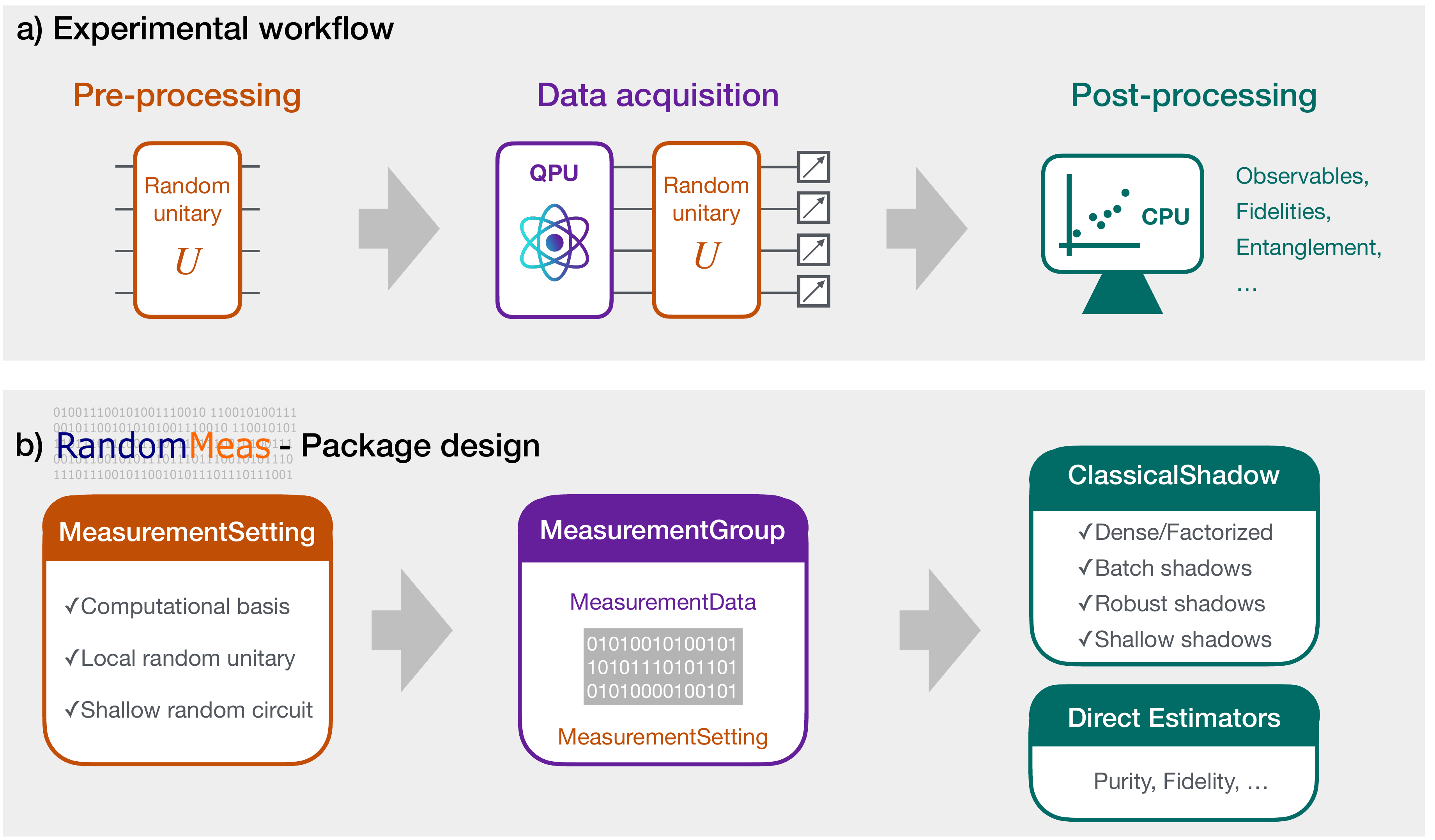}
\end{center}
 \caption{{\it Experimental RM workflow  and corresponding organization of the package \texttt{RandomMeas.jl}. \label{fig:setup}} (a) Pre-processing consists in sampling random unitaries, specifying randomized measurement settings, from a prescribed ensemble. During data acquisition random unitaries are implemented on the state (or process) of interest and projective measurements are performed. Classical post-processing predicts a wide range of quantum state (or quantum process) properties. (b) 
 Within \texttt{RandomMeas.jl}, pre-processing, data acquisition and post-processing are separated in three independent sets of data structures. Data acquisition can be performed with a quantum device (quantum processing unit, QPU), or, alternatively, simulated classically on a CPU. For the latter, \texttt{RandomMeas.jl} provides  tensor network routines based on \texttt{ITensors.jl}.}
\end{figure*}

The package is modular by design. Its public API revolves around three core object types—\emph{measurement settings}, \emph{measurement data}, and \emph{classical shadows}—that can be flexibly combined to implement and extend RM protocols. Advanced features such as robust shadows~\cite{chen_robust_2021,koh_classical_2022,vitale_robust_2024}, shallow shadows~\cite{hu_classical_2023,akhtar_scalable_2023,bertoni_shallow_2024}, batch shadow estimation~\cite{rath_entanglement_2023}, and common randomized measurements~\cite{vermersch_enhanced_2024} can be enabled via simple keyword switches.

\paragraph{Comparison to existing software.}

Several software libraries support aspects of randomized measurements (RM), each serving different goals and user groups. 
The \texttt{povm-toolbox}~\cite{fischer2024dual} for Qiskit enables local and global POVMs, including arbitrary informationally complete measurements, and provides basic classical shadow routines. 
PennyLane offers a tutorial-level implementation~\cite{RoelandWiersema2021} and a \texttt{ClassicalShadow} class for random Pauli measurements, integrated into its broader quantum machine-learning framework and suited for small- to moderate-scale demonstrations. 
Mitiq incorporates primitives for robust classical shadows with calibration-based error mitigation and supports multiple quantum backends, making it a flexible platform for exploring error-mitigated RM, though simulations are generally restricted to \(\lesssim 20\) qubits \cite{larose2022mitiq}. 
Other toolkits, such as \texttt{TensorCircuit} (a general-purpose tensor-network simulator) \cite{Zhang2023tensorcircuit} and \texttt{Paddle Quantum} (educational demonstrations) \cite{Paddlequantum}, include basic classical shadow capabilities within broader simulation or teaching-focused contexts.

\change{
In the small-system regime where state-vector simulation is feasible (typically \(\lesssim 20\) qubits), these packages overlap in functionality for standard Pauli classical shadows (and closely related local measurement schemes), and performance depends strongly on the backend, batching strategy, and estimator workload.
Beyond this, \texttt{RandomMeas.jl} also supports a broader class of RM protocols, including arbitrary local random-unitary (RU) ensembles, global RU measurements, and circuit-based shallow shadows.
\texttt{RandomMeas.jl} is designed to support the \emph{full RM workflow in a scalable, modular, and high-performance setting}.
It provides tensor-network-based simulation and post-processing for RM protocols, enabling the simulation and analysis of large-scale systems beyond the reach of state-vector methods.
It supports multiple classical shadow variants—including error-mitigated robust, shallow, and batch shadows—and provides efficient estimators for both linear and polynomial functionals with built-in statistical error quantification via resampling techniques.
\texttt{RandomMeas.jl} is designed for seamless integration into both actual QPU experiments and large-scale classical simulations.
Its emphasis on modularity, composability, and reproducibility makes it uniquely suited for developing, testing, and deploying advanced RM protocols in practice.}

\change{
To facilitate practical tool selection, we provide benchmark scripts in the repository (for representative RM workloads) that record the execution environment (Julia version and package versions) and can be rerun across machines and problem parameters. In particular, \lstinline{benchmarks/scaling\_benchmarks.ipynb} and \lstinline{benchmarks/example\_benchmarks.jl} cover representative simulation and post-processing workloads (dense vs.\ factorized shadow construction, local Pauli expectation values, and reduced-density-matrix moments) and report both run time and memory footprints.}

\paragraph{Article overview.}

This article is organized as follows.  
Section~\ref{sec:RM} introduces the core concepts behind randomized measurements and outlines the software architecture.  
Section~\ref{sec:basic} walks through a complete RM workflow, from setting generation to observable estimation.  
Subsequent sections present advanced features, including robust and shallow shadows, batch shadow estimation, and polynomial functional estimation, illustrated through a series of curated \textsc{Jupyter} notebooks provided in the package's \lstinline{examples} folder. \change{These notebooks are described in more detail in appendix \ref{app:jupyter}.}

\smallskip
\noindent
\texttt{RandomMeas.jl} is released under the Apache 2.0 license and
available at \url{https://github.com/bvermersch/RandomMeas.jl}. The documentation can be found here \url{https://bvermersch.github.io/RandomMeas.jl/dev/}. The specific version corresponding to this paper is tagged with \href{https://github.com/bvermersch/RandomMeas.jl/releases/tag/v0.3.1_quantum_publication}{\texttt{v0.3.1\_quantum\_publication}} and archived on Zenodo as Ref.~\cite{vermersch_elben_randommeas_zenodo}.

\section{Randomized-measurement workflow and package architecture}
\label{sec:RM}

\subsection{Randomized-measurement workflow}

Randomized-measurement (RM) protocols allow one to access many
observables and entropic quantities of a quantum state
$\rho$ while requiring only projective measurements in
random bases~\cite{elben_randomized_2022}.
A convenient way to describe an RM
experiment, and, correspondingly, the design philosophy of
\texttt{RandomMeas.jl}, is to split the procedure into three stages
(see Fig.~\ref{fig:setup}):

\medskip
\noindent\textbf{(i) Pre-processing.}\;
Measurement settings are specified by
a random unitary $U_{\mathrm{set}}$ sampled from a prescribed ensemble.
The most common choice is a product of single-qubit rotations,
$U_1\otimes\dots\otimes U_N$,
but shallow multi-qubit circuits can
be advantageous in specific scenarios.
The total number of sampled settings is denoted
$N_{\mathrm U}$.

\smallskip
\noindent\textbf{(ii) Data acquisition.}\;
On the quantum processor each unitary~$U_{\mathrm{set}}$
is applied, followed by a computational-basis measurement.
Repeating the measurement $N_{\mathrm M}$ times produces
$N_{\mathrm U}\times N_{\mathrm M}$ bit strings
$\mathbf s\in\{0,1\}^N$.
The collection of outcomes and employed settings is called a \emph{measurement group}.

\smallskip
\noindent\textbf{(iii) Post-processing.}\;
The recorded bit strings, together with the employed measurement settings, are processed classically to estimate properties of the quantum state of interest.
 To this end, we can use the classical shadow formalism to define classical snapshots $\hat{\rho}$ that can, for instance, be used to estimate expectation values of observables $O$ via $\hat{O}=\text{Tr}[O \hat{\rho}]$. The precise definition of classical shadows depends on the ensemble of measurement settings, typically one chooses them to be  unbiased estimators of the density matrix, i.e., $\mathbb{E}[\hat{\rho}]=\rho$, where $\mathbb{E}$ is the expectation value over both the random ensemble of measurement settings and the quantum mechanical average. With local measurement settings where each single qubit unitary $U_i$ is sampled from a unitary 3-design \cite{elben_randomized_2022}.  Then, the classical shadow is defined as~\cite{huang_predicting_2020}  
\begin{align}
   \hat{\rho}
  \;=\;
  \bigotimes_{i=1}^N
  \bigl(
    3\,U_i^\dag\ket{s_i}\!\bra{s_i}U_i - \openone
  \bigr).   
\end{align}
where $\mathbf{s}=(s_1,\dots,s_N)$ is the measurement outcomes of the computational-basis measurement after applying $U$.

In special cases, for instance purity estimation
$\operatorname{tr}(\rho^2)$~\cite{elben_renyi_2018} or cross-entropy fidelity estimation \cite{boixocharacterizing2018}, one can bypass the
shadow step and apply analytic estimators directly to the measurement
group.

\subsection{Package design}

\texttt{RandomMeas.jl} mirrors the conceptual workflow  directly in its software architecture by providing three families
of data structures:

\paragraph{Measurement settings.}
The abstract type \texttt{MeasurementSetting} represents a single
measurement setting.  Its most frequently used subclass,
\texttt{LocalMeasurementSetting}, covers the case of tensor–product
unitaries, with concrete implementations
\texttt{LocalUnitaryMeasurementSetting} (independently sampled
single–qubit rotation per site from the Haar measure on $U(2)$ or $X,Y,Z$ rotations) and
\texttt{ComputationalBasisMeasurementSetting} (the identity on every
site).  The package also provides
\texttt{ShallowUnitaryMeasurementSetting}, which stores an ordered list
of one- and two-qubit gates that realize a shallow depth circuit to implement the measurement setting.

\paragraph{Measurement data.}
A \texttt{MeasurementData} instance holds the bit strings obtained from
projective measurements performed in a single setting, together with the
setting itself.  Several such objects, collected over
$N_{\mathrm U}$ different settings, are gathered into a
\texttt{MeasurementGroup}.  The library offers simulators
that, given a tensor-network representation of a quantum state~$\rho$,
simulate the entire acquisition process and return a
\texttt{MeasurementGroup} identical in form to experimental data.

\paragraph{Classical shadows.}
Post-processing routines evaluate \texttt{MeasurementGroup} objects, and evolve  around the abstract type
\texttt{AbstractShadow}.  Three concrete  formats for classical shadows are available.
\texttt{FactorizedShadow} is memory–efficient and scales to essentially
arbitrary $N$ by storing vectors of $2 \times 2$ single-qubit classical shadows per
site; \change{\texttt{DenseShadow} stores a classical shadow as a full $2^{N}\times 2^{N}$ object. Since memory and basic linear-algebra costs scale exponentially in $N$ (i.e.\ with $4^{N}$ matrix entries), this representation is primarily intended for small systems or small subsystems (typically $N$ or $N_A \sim 10$--$12$, depending on hardware and how many shadows/batches are kept in memory), where it can be faster than factorized shadows}; finally
\texttt{ShallowShadow} is tailored to measurement settings generated by
shallow circuits.  In the context of dense representations, \texttt{MeasurementProbability}
objects contain, for each setting, the empirical Born distribution (as a $2^N$ dimensional normalized vector)
extracted either from the raw bit strings (\texttt{MeasurementData}) or via simulations.  A suite of estimator functions then
acts on classical shadows or directly on \texttt{MeasurementGroup} data structures to compute fidelities, purities,
overlaps, Rényi entropies, \emph{etc.}.

\medskip
All tensor algebra is delegated to the highly optimised
\texttt{ITensors.jl} library~\cite{fishman_ITensor_2022}, which
enables both efficient classical simulation of randomized measurements in large quantum systems and convenient manipulation of high-rank tensors that arise
during post-processing.

\subsection{Installing and usage}
\medskip

The \RM package can be easily installed using the Julia package manager. To install, simply execute the following command in the Julia REPL:
\begin{lstlisting}
import Pkg
Pkg.add(RandomMeas)
\end{lstlisting} To use \RM, one can then import the package and check its version.
\begin{lstlisting}
using RandomMeas
RandomMeas.version()
\end{lstlisting}

\section{Basic usage}\label{sec:basic}

\begin{figure}[t]           
  \centering
  \includegraphics[width=\columnwidth]{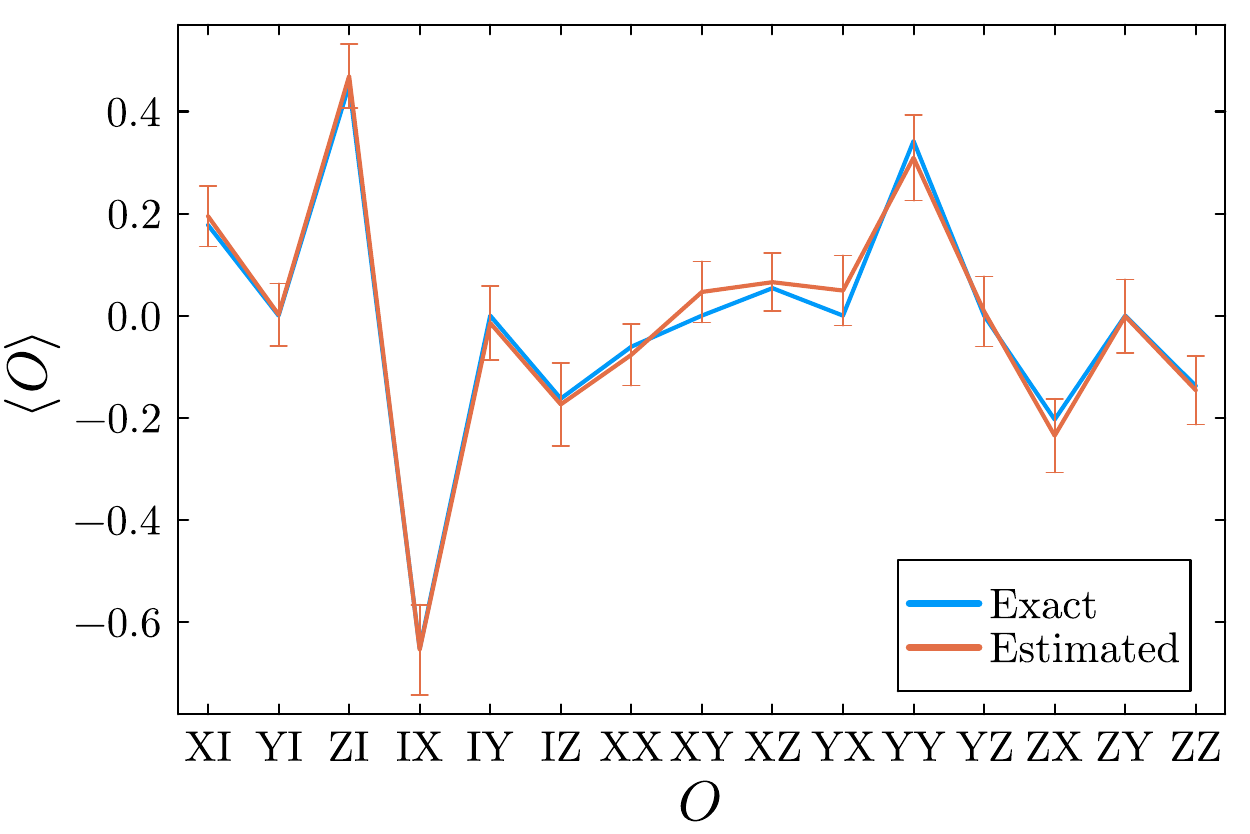}
  \caption{{\it Estimating expectation values with \RM.}
  Using a random MPS of $N=50$ qubits, bond dimension 2, and randomized measurements with $N_U=200$, $N_M=100$, we extract all Pauli observables with support on sites $1$ or $4$, as described in Sec.~\ref{sec:basic}. The errors bars are obtained following Sec.~\ref{sec:stats}, and correspond throughout this manuscript to $\pm 2$ estimated standard deviations.
  \label{fig:expectation}}
\end{figure}

This section (and the corresponding notebook) walks through one complete workflow of the package – from sampling random measurement settings, through collecting  measurement data, to building classical shadows and  estimating observables and non-linear functionals. In the last subsection, we show how to estimate the statistical accuracy via the Jackknife procedure.

\subsection{Data acquisition}

\paragraph{Randomized measurement settings.}
We start by importing the package and define the (total) system size of interest $N$. We now sample randomized measurement settings. Here, we parametrize them through single‑qubit (local) random unitaries which are drawn for each qubit independently from the Haar ensemble on $U(2)$. We draw $N_U = 200$ settings. 
\begin{lstlisting}
N = 50 # Number of qubits
NU = 200 # Number of measurement settings
measurement_settings = 
   [LocalUnitaryMeasurementSetting(N) for _ in 1:NU]
\end{lstlisting}

\change{\paragraph{Interfacing \RM with quantum experiments.}
As a next step, randomized measurements can be performed on actual quantum hardware.
On the one hand, \RM provides routines for exporting measurement settings. This includes interfaces to \texttt{qiskit} and,
for \texttt{LocalUnitaryMeasurementSetting}, an OpenQASM export routine (QASM~2.0). These workflows are demonstrated in the
example notebooks (see also App.~\ref{app:jupyter}) and illustrate how to deploy the generated measurement circuits using external software frameworks.
On the other hand, \texttt{qiskit} (or any other native quantum programming language) can be used to directly sample \(N_U\)
random measurement settings and express them as parameterized quantum circuits (see the dedicated \texttt{qiskit} example
notebooks, App.~\ref{app:jupyter}).}

\change{After performing randomized measurements on quantum hardware, the measured bitstrings and the corresponding random
unitaries can be post-processed using \RM, for instance by constructing a \texttt{MeasurementGroup} as:}
\begin{lstlisting}
basis_transfo = npzread("data/qiskit_u.npy")
meas = 2 .- npzread("data/qiskit_data.npy")
measurement_group = MeasurementGroup(meas, basis_transfo)
\end{lstlisting}
\change{as detailed in the example notebooks (see also App.~\ref{app:jupyter}).}

\paragraph{Classical simulation of randomized measurements.}
Alternatively, one can simulate the entire RM protocol. Here, we choose a weakly entangled  matrix product state (MPS) $\ket{\psi}$ with bond dimension 2.
\begin{lstlisting}
site_indices = siteinds("Qubit",N)
ψ = random_mps(site_indices, linkdims=2) # Quantum state of interest
\end{lstlisting}
For each setting, we perform $N_M = 100$ projective measurements, yielding a ``MeasurementGroup`` container that stores all measurement outcomes and measurement settings used.
\begin{lstlisting}
NM=100 # Number of projective measurements per setting
measurement_group = MeasurementGroup(ψ ,measurement_settings,NM)
\end{lstlisting}
\change{This constructor also supports input states provided either as an MPS (pure state) or as an MPO (density operator).
The optional keyword \texttt{mode} selects tensor-network sampling (default) or dense-array sampling (when feasible).}

\subsection{Classical shadows and estimation of expectation values}\label{sec:expectation}

\change{\paragraph{Building classical shadows}
The first steps of post-processing consist usually in building classical shadows~\cite{huang_predicting_2020}.
Here, we first use "factorized" classical shadows that are memory efficient objects (stored as N x 2 x 2 arrays) suitable for large system sizes.}
\begin{lstlisting}
classical_shadows = get_factorized_shadows(measurement_group)
\end{lstlisting}

\paragraph{Expectation values.}

For estimating expectation values, we construct an observable
$$
O = Z_1 \otimes I_2 \otimes I_3 \otimes X_4 \otimes I_{5} \otimes \dots \otimes I_N.
$$
\begin{lstlisting}
ops = ["I" for _ in 1:N]
ops[1] = "Z"
ops[4] = "X"
O=MPO(site_indices,ops)
\end{lstlisting}
We compute the shadow estimate $\widehat{\langle O\rangle}$ and its standard error, and the exact value $\langle O\rangle = \langle\psi\lvert O \rvert\psi\rangle$ . 
\begin{lstlisting}
estimate_val = get_expect_shadow(O,classical_shadows)
exact_val = inner(ψ',O,ψ)
\end{lstlisting}

\paragraph{Dense shadows for small subsystems.}

When the observable acts only on a few qubits it is efficient to
\emph{restrict} the data to that subsystem and keep each shadow in its
full matrix form.  For a subsystem of $N_A$ qubits a dense classical shadow is stored as a full $2^{N_A}\times 2^{N_A}$ object (i.e.\ $4^{N_A}$ entries). \change{ In practice, dense shadows are typically advantageous for small $N_A$ (roughly $N_A\sim 10$--$12$ on a workstation, depending on batching and how many shadows are stored/processed at once); for larger $N_A$, factorized shadows are usually preferable due to the exponential memory \emph{and} arithmetic cost. A reproducible timing/memory comparison underlying this guideline is provided in \lstinline{benchmarks/scaling\_benchmarks.ipynb}.}

In our example the operator \(O\) has support on qubits \(1\) and \(4\)
only.  The three-step workflow is therefore: reduce the measurement data to the sites \(\{1,4\}\); build dense shadows for that two-qubit subsystem; estimate \(\langle O\rangle\) with those shadows.

\begin{lstlisting}
reduced_group  = reduce_to_subsystem(measurement_group, [1, 4])
dense_shadows = get_dense_shadows(reduced_group)
estimate_val  = get_expect_shadow(reduce_to_subsystem(O,[1,4]), dense_shadows)
\end{lstlisting}
Both routes - factorized and dense - yield the same estimated value for
\(\langle O\rangle\);
the dense-shadow routines are merely faster on smaller
(sub-) systems.

An example of estimation of multiple Pauli strings from the same RM dataset is shown in Fig.~\ref{fig:expectation}.

\subsection{Estimating non-linear functionals}

Many physically relevant quantities—such as purities
\(p_2=\operatorname{tr}(\rho^{\,2})\) or higher‐order Rényi entropies—are
polynomial functionals of the density matrix. Classical shadows provide a natural
way to estimate such polynomial functionals via the so-called
\emph{U-statistics} formalism \cite{huang_predicting_2020,elben_mixed-state_2020}: to estimate, for instance,
$\operatorname{tr}(\rho^{\,k})$, one replaces each factor in the trace
product by a statistically independent classical shadow (obtained from
independently sampled random unitaries) and averages over all distinct
combinations~\cite{huang_predicting_2020,elben_mixed-state_2020}. For $N_M=1$, this yields
\begin{align}
\label{eq:Ustat}
\widehat{p}_k \;=\;
\frac{(N_U-k)!}{N_U!}
\sum_{\substack{j_1,\dots,j_k\\\text{all distinct}}}
\operatorname{tr}\!\bigl[\hat\rho_{j_1}\hat\rho_{j_2}\dotsm\hat\rho_{j_k}\bigr],
\end{align}
which is an unbiased estimator of the $k$-th moment
\(p_k=\operatorname{tr}(\rho^{\,k})\) for $k>1$ integer.

For general $N_M>1$, the combinatorics of the U-statistic become more
involved, but \texttt{RandomMeas.jl} handles this general case. Thus, users can directly call
\lstinline{get_trace_moments(shadows, collect(2:5))} for arbitrary $N_U$
and $N_M$ without worrying about the underlying formulas.

Evaluating the U-statistic exactly 
scales as $\frac{(N_U-k)!}{N_U!}N_M^k$. 
It therefore becomes quickly impractical when \(N_U\), \(N_M\) and \(k\) grow.
A crucial practical simplification is obtained by introducing
\emph{batch shadows} \cite{rath_entanglement_2023}. Here the $N_U$ unitaries are partitioned into $N_B$
disjoint batches of size $N_U/N_B$, and one averages within each
batch (including over all $N_M$ outcomes), defining
\(
\bar\rho_b=\tfrac{1}{(N_U/N_B) N_M}\sum_{j\in b}\sum_{l=1}^{N_M}\hat\rho_{j,l}
\) where $\hat\rho_{j,l}$ is the classical shadow constructed from measurement $l$ after the application of the random unitary $j$.
The resulting $N_B$ batch shadows are then combined according to the same
U-statistic formula~\eqref{eq:Ustat}, but with $N_U$ replaced by $N_B$,
reducing the number of terms to $\tfrac{(N_U-k)!}{N_U!}$ while retaining the
leading $1/N_U$ variance in the practically relevant regime
$N_B\ll N_U$~\cite{rath_entanglement_2023}.

In \texttt{RandomMeas.jl}, this batching workflow is accomplished in two
lines:
\begin{lstlisting}
batch_shadows = get_dense_shadows(measurement_data; n_ru_batches = 8)
moments = get_trace_moments(batch_shadows, collect(2:5))
\end{lstlisting}
The first call forms $N_B=8$ batch-averaged dense shadows, while the
second evaluates the reduced U-statistic on them.

Another simplification of the post-processing consists
in using estimation formulas that do not require building classical shadows but work directly with the obtained measurement data. This is the case in particular  for the purity~\cite{elben_renyi_2018,brydges_probing_2019}, and cross-platform fidelity~\cite{elben_cross-platform_2020}. 
Our package provides the two functions \texttt{get\_purity}, and \texttt{get\_overlap} that take \texttt{MeasurementGroup} objects as inputs and return the purity and the state overlap $\mathrm{tr}(\rho_1 \rho_2)$.

\begin{figure}[t]
  \centering
  \includegraphics[width=\columnwidth]{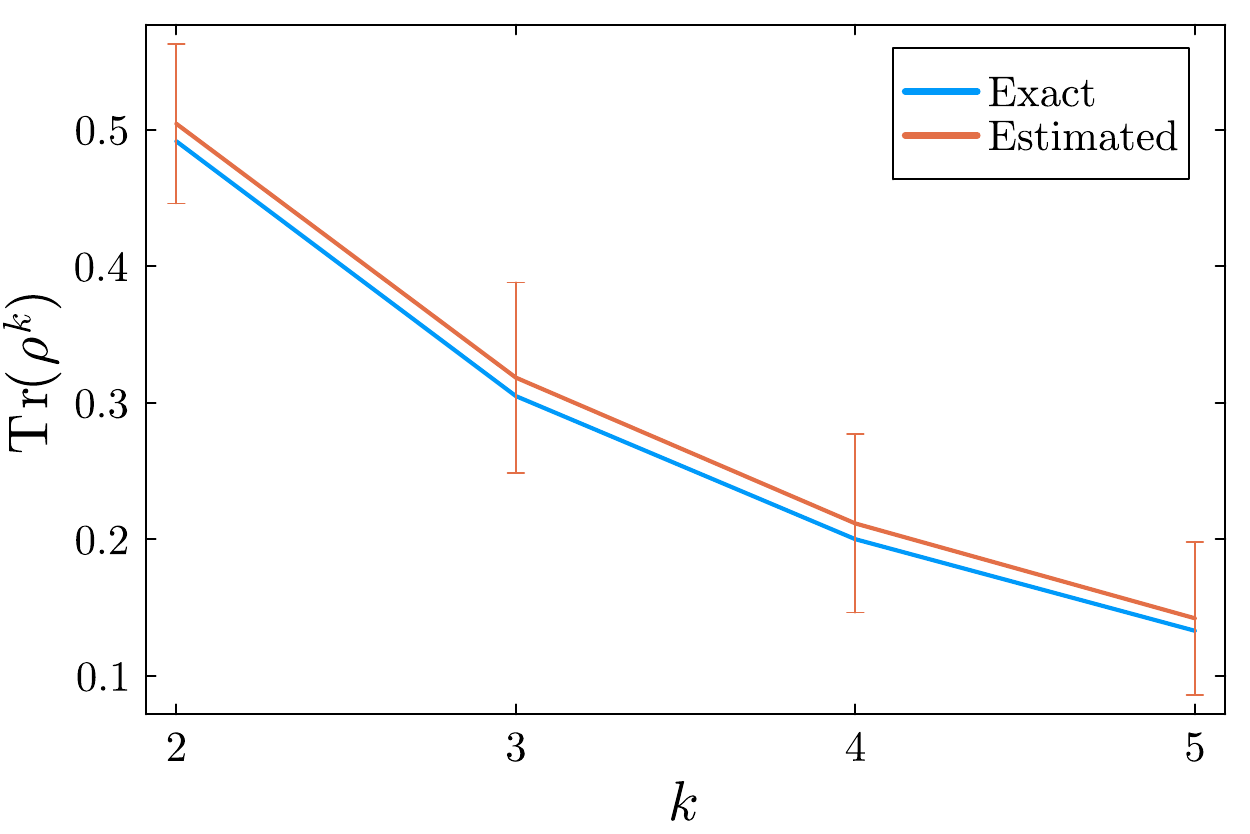}
  \caption{Estimation of trace moments $p_k=\mathrm{tr}(\rho^k)$ using the same RM data as in Fig.~\ref{fig:expectation}, and using $8$ batch shadows.
  \label{fig:moments}}
\end{figure}

\subsection{Quantifying statistical accuracy}
\label{sec:stats}

Randomized–measurement protocols infer physical quantities from finite
samples of random bases and finite numbers of projective read-outs.  Any
numerical estimate produced by \RM must therefore be accompanied by an
\emph{uncertainty} that reflects two independent sources of
fluctuations, \emph{ensemble noise} arising from the limited number $N_{\mathrm U}$ of sampled measurement settings; and \emph{shot noise} originating from the finite number
      $N_{\mathrm M}$ of projective measurements  per setting.
      
\paragraph{Linear functionals.}
For estimates that are \emph{linear} in the state---e.g.\ an observable
expectation value $\operatorname{tr}(O\rho)$---the estimator is the
sample mean of $N_B$ (batch-)shadow evaluations.  The routine
\lstinline{get_expect_shadow} (and analogues) accepts the keyword
\lstinline{compute_sem=true} and returns the value together with its
standard error
\(
\sigma = \sqrt{\operatorname{Var}[\,\hat O\,]/N_B}\,
\).

\paragraph{Polynomial functionals.}
Non-linear quantities such as purity
$\operatorname{tr}(\rho^2)$ of a quantum state $\rho$ or higher trace moments admit unbiased
\emph{U-statistic} estimators but lack simple analytic variance
formulae.  \RM therefore implements the leave-one-out
\emph{jackknife} \cite{wu1986jackknife} to estimate the variance of the corresponding estimate:
\begin{enumerate}
  \item From $N_B$ (batch) shadows compute the U-statistic $\hat\theta$.
  \item For each $i=1,\dots,N_B$ omit shadow~$i$ and recompute
        $\hat\theta_{(i)}$ from the remaining $N_B-1$ shadows.
  \item The jackknife bias‐corrected point estimate is
        \(\theta_{\text{JK}} = N_B\hat\theta - (N_B-1)\,\overline{\hat\theta}_{(\cdot)}\),
        with variance
        \(\sigma_{\text{JK}}^2 = \tfrac{N_B-1}{N_B}\sum_i
          (\hat\theta_{(i)}-\overline{\hat\theta}_{(\cdot)})^{2}\).
\end{enumerate}
A direct implementation requires $\mathcal O(N_B)$ evaluations of the
costly U-statistic.  \RM avoids this overhead by caching all terms that
enter the full estimator and re-aggregating them to obtain every
$\hat\theta_{(i)}$ essentially for free; the run time therefore remains
dominated by the single computation of the full U-statistic. It also provides functionality to estimate the full covariance matrix of multiple trace moments via jackknife.

\paragraph{Practical guidelines.}
Reliable uncertainties require a minimum of roughly ten (batch) shadows
($N_B\gtrsim10$) \cite{rath_entanglement_2023}.  Users are encouraged to check convergence of both the
point estimate and the jackknife error bar as a function of~$N_B$.
\change{We note that bootstrap resampling is another natural option for uncertainty quantification, but it is not currently implemented in \RM; we focus on jackknife as a deterministic and computationally efficient default.}

All post-processing routines expose the keywords
\lstinline{compute_sem} / \lstinline{compute_cov} to activate these
statistical diagnostics; by default only the point estimate is
returned.

\section{Advanced usage of the package}

In the section, we present examples of advanced usage of the toolbox.
First we present two illustrative examples for performing classical shadow tomography: In presence of measurement noise (Sec.~\ref{sec:robust}), and using shallow quantum circuits as measurements settings (Sec.~\ref{sec:shallow}).
Then, in Sec.~\ref{sec:jupyter}, we present a list of other possible advance uses of the toolbox, and connect them to jupyter notebooks that are available online (see also App.~\ref{app:jupyter}).

\subsection{Robust shadow tomography}\label{sec:robust}

\begin{figure}[t]        
  \centering
\includegraphics[width=\columnwidth]{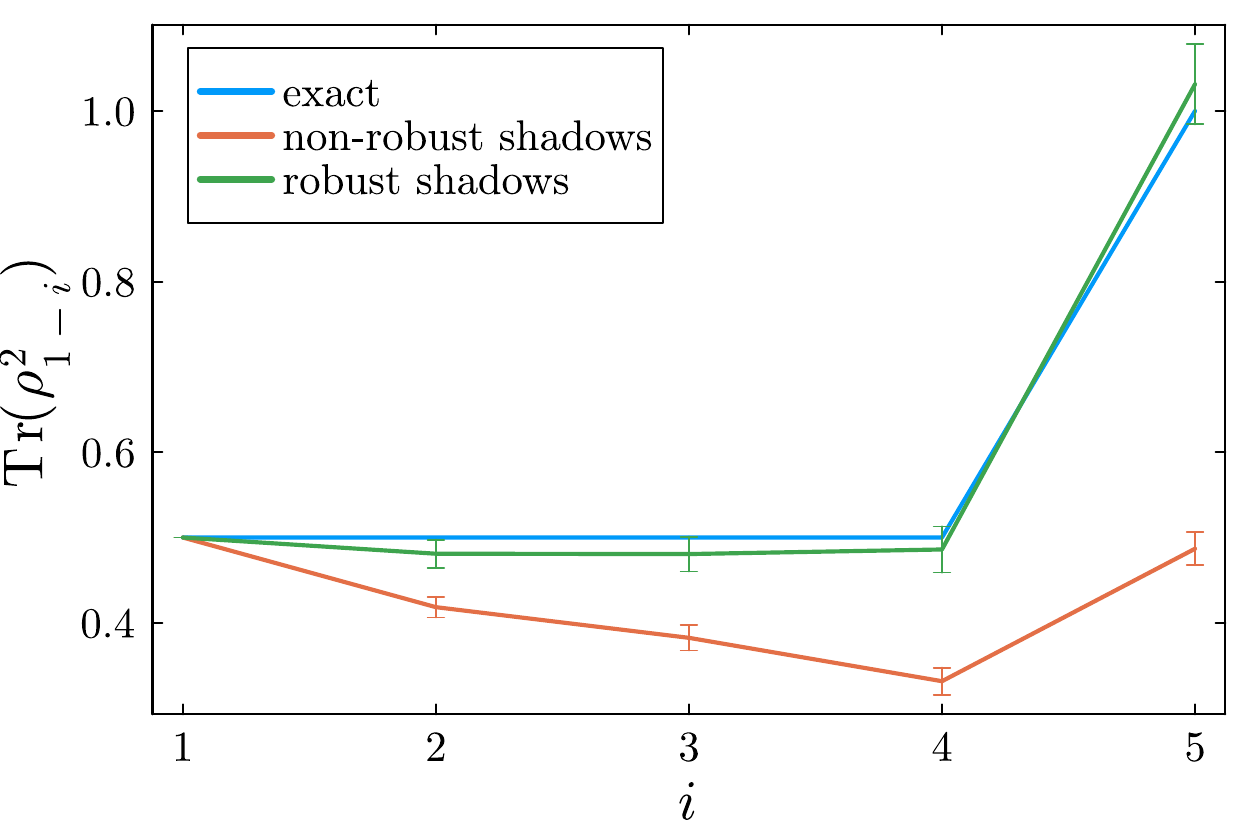}
  \caption{Robust shadow estimation of the purity of (sub-)systems of a $5$ qubit GHZ state as a function of the subsystem size $i$.  The measurement is affected by local depolarization noise of random strength of mean value $0.1$ and standard deviation $0.02$. We used  $N_U=2000$ unitaries  for the calibration experiment and for the measurement ($N_M=100$ for both).}
  \label{fig:robust}
\end{figure}

\noindent
Randomized–measurement protocols are appealing precisely because they are
\emph{hardware friendly}: in their simplest version, only local basis rotations and computational basis measurements are needed.  In practice, however, during the
basis rotations, read-out crosstalk and errors 
degrade the quality of the recorded bit strings.
Classical shadow estimators that neglect those imperfections can be
severely biased.

\paragraph{Twirling and effective noise.}
A key observation---formalised in
Refs.\,\cite{chen_robust_2021,koh_classical_2022}---is
that the ensemble average over random unitaries \emph{twirls} the native
noise channel into a simple, diagonal form\footnote{Under the assumption of gate-independent, time-stationary and Markovian noise.}.
For local single-qubit rotations the effective noise on each qubit is
fully characterized by a single \emph{depolarization parameter}
$\,G_i\,$ (or, equivalently, an error probability $p_i$).  
If these parameters are known, one can
\emph{undo} the bias at the level of the shadow map\footnote{%
All mitigation is performed \textit{a posteriori} at the expense of a possibly enlarged sample complexity: no additional pulses are
inserted into the experimental sequence.} and recover an 
unbiased estimator.

\paragraph{Calibration from a reference experiment.}
To measure \emph{depolarization parameters}, we  start from a known product state $\ket{\psi_0}$ (typically $\ket{\psi_0}=\ket{\mathbf{0}}$) and acquire a measurement group \texttt{calibration\_group}, which will be used to build robust shadows for a subsequent experiment performed on a unknown quantum state $\rho$.

Based on this data, \RM first constructs a calibration vector $G$~\cite{vitale_robust_2024} containing he per-qubit depolarization parameters, which can be then passed to any shadow-construction routine for the post-processing of the actual \texttt{measurement\_data}.
(\lstinline{get_factorized_shadows}, \lstinline{get_dense_shadows}, \dots). 

\begin{lstlisting}
G_e = get_calibration_vector(ψ0,calibration_group)
robust_shadows = get_factorized_shadows(measurement_data; G)
\end{lstlisting}

Because the correction is inserted at the earliest possible point (the
definition of the classical shadow itself) \texttt{RandomMeas.jl} offers
robust versions of \emph{every} estimator implemented in the library.

\paragraph{Visual illustration.}
Figure~\ref{fig:robust} presents the standard and the robust
estimators for the purity of a noisy GHZ state, as in the experimental work~\cite{vitale_robust_2024}, see also the corresponding jupyter notebook. While standard, i.e.\ non-robust, classical shadows, underestimate the values of the purity, the robust version  rescales the estimator to remove the effect of measurement errors.

\subsection{Shallow shadows}\label{sec:shallow}

Classical shadows constructed from local measurement settings are hardware-friendly and efficient for estimating local observables, but may yield large statistical errors when estimating high-weight operators\footnote{The number of measurements scale exponentially with the weight of the observables of interest \cite{huang_predicting_2020}.}.
An alternative, known as \emph{shallow shadows}, replaces product random unitaries with low-depth random quantum circuits. 
This approach can substantially reduce estimator variance for high-weight observables~\cite{hu_classical_2023,akhtar_scalable_2023,ippoliti2023operator,bertoni_shallow_2024,hu_demonstration_2025}.

\paragraph{Learning the inverse shadow map.}
In \RM shallow shadows are implemented using uni-dimensional random quantum circuits composed of single-qubit and nearest-neighbor two-qubit gates of fixed depth. In contrast to local-unitary schemes, the corresponding shadow map is not analytically known and must be learned numerically from training data. This is done during a training phase using a vector of \texttt{ShallowMeasurementSetting} instances.
\begin{lstlisting}
NU_training = 2000 # Number of unitaries
N = 10
site_indices_i = siteinds("Qubit", N;addtags="input")
site_indices_o = siteinds("Qubit", N;addtags="output")
depth = 2
settings = [ShallowUnitaryMeasurementSetting(N,depth;site_indices=site_indices_o) for _ in 1:NU_training];
\end{lstlisting}
Here we use two instances of site indices, \texttt{site\_indices\_i}, \texttt{site\_indices\_o}, in order to describe the action of quantum channels on input and output states, respectively.

From the resulting data, the effective measurement channel is reconstructed as
\begin{equation}
\mathcal{M}(\rho) = \mathbb{E}_U \left[ \sum_\mathbf{s} \bra{\mathbf{s}} U \rho U^\dagger \ket{\mathbf{s}} \, U^\dagger \ket{\mathbf{s}} \bra{\mathbf{s}} U \right],
\end{equation}
where the average is taken over the shallow circuit ensemble.

Internally, \(\mathcal{M}\) is represented as a depolarizing channel parametrized by a real-valued MPS. Using the sets of $N_U$ unitaries defined through the variable \texttt{settings}, this MPS is obtained trough a tensor-network fitting procedure~\cite{baumgratz_scalable_2013}, using the numerically evaluated Born probabilities generated from a fixed state $\rho_0=\ket{0}\bra{0}$.
In a second step, we compute the inverse map  \(\mathcal{M}^{-1}\) via automatic differentiation. More theoretical details are given in the corresponding Jupyter notebook. 

\paragraph{Observable estimation.}
Once the inverse map is learned, it can be used to construct classical shadows from randomized measurements on unknown quantum states. Here, we simulate these randomized measurements classically using the built-in functionality of \RM. A new measurement group is collected using the same circuit depth:
\begin{lstlisting}
measurement_group = MeasurementGroup(ψ,NU,NM;setting_type=ShallowUnitaryMeasurementSetting,depth=depth, mode=Dense);
\end{lstlisting}
Each measurement outcome \(\mathbf{s}\), obtained from a shallow circuit \(U\), is then mapped to a classical shadow via
\begin{equation}
\hat{\rho} = \mathcal{M}^{-1}(U^\dagger \ket{\mathbf{s}}\bra{\mathbf{s}} U),
\end{equation}
enabling unbiased estimation of expectation values. For instance, an observable \(O_j\) can be evaluated as
\begin{lstlisting}
shadow = get_shallow_shadows(measurement_group, inverse_shallow_map, site_indices_i, site_indices_o)
estimate = real(get_expect_shadow(observable[j], shadow))
\end{lstlisting}
This procedure supports scalable estimation of many observables, and can significantly reduce variance for non-local observables compared to standard local-measurement shadows \cite{ippoliti2023operator}.

\subsection{Benchmarking of quantum devices}

Randomized measurement protocols also enable practical and scalable benchmarking of quantum processors. \RM includes built-in support for two widely used protocols: cross-entropy benchmarking (XEB) \cite{boixo2018characterization,arute_quantum_2019} for comparing a (noisy) QPU against an ideal classical simulation and cross-platform fidelity estimation for comparing two QPUs \cite{elben_cross-platform_2020}. These routines exploit efficient tensor-network simulation to benchmark realistic circuits and noisy dynamics at scale.

\paragraph{Cross-entropy benchmarking (XEB) and self-XEB.}
Cross-entropy benchmarking~\cite{boixo2018characterization,arute_quantum_2019} quantifies hardware fidelity by comparing the (empirical) probabilities of experimentally measured bitstrings to the ideal output probabilities \(p(\mathbf{s}) = |\langle \mathbf{s} | U | \mathbf{0}\rangle|^2\) of a random quantum circuit \(U\). Given a dataset of bitstrings \(\{\mathbf{s}\}\), the XEB fidelity is computed as
\begin{equation}
\mathcal{F}_{\text{XEB}} = 2^N \cdot \mathbb{E}_{\mathbf{s} \sim \text{exp.}} [p(\mathbf{s})] - 1,
\end{equation}
where the expectation is over experimental bitstrings and \(p(\mathbf{s})\) is computed for each experimental measurement outcome $\mathbf{s}$ using a classical simulation of the ideal circuit.
For shallow circuits, the output distribution can deviate significantly from the Porter–Thomas form, leading to a biased XEB estimator. 
The \emph{self–XEB} protocol~\cite{andersen_thermalization_2025} (see also Refs.~\cite{Mark_2023,Shaw_2024}) corrects for this by normalizing with probabilities obtained directly from the reference circuit, yielding
\[
\mathcal{F}_{\mathrm{self-XEB}} = \frac{\mathbb{E}_{\mathbf{s} \sim \text{exp.}} [p(\mathbf{s})] - \overline{p}}{\mathbb{E}_{\mathbf{s} \sim \text{ref.}} [p(\mathbf{s})] - \overline{p}},
\]
where \(\overline{p} = 1/2^N\) is the uniform baseline.

Both XEB and self–XEB are supported directly from \lstinline{MeasurementData} objects.  
\begin{lstlisting}
XEB = get_XEB(ψ, measurement_data)     # standard XEB
selfXEB = get_selfXEB(ψ)               # self-XEB reference
XEB_corrected = XEB / selfXEB          # bias-corrected estimate
\end{lstlisting}
Here, \lstinline{get_XEB} computes the standard XEB estimator with measurement data from computational basis measurements on a noisy version of $\psi$. The second routine evaluates the self-XEB normalization for $\psi$. 
In this context, the package can simulate ideal and noisy dynamics using matrix–product–state and quantum–trajectory methods~\cite{jaschke_one-dimensional_2018}.


\subsection{Jupyter notebooks for advanced usage}\label{sec:jupyter}

To illustrate the use of \texttt{RandomMeas.jl} in realistic scenarios, we provide a collection of 14 Jupyter notebooks, listed in App.~\ref{app:jupyter}.   
They cover a broad range of protocols and use cases, from classical–shadow tomography (including robust and shallow–circuit variants) and quantum–device benchmarking (e.g., cross–entropy benchmarking, cross–platform fidelity estimation) to entanglement–entropy measurements and topological–order detection.  
Several notebooks reproduce or reanalyse experimental data, such as the ion–trap experiment of Ref.~\cite{brydges_probing_2019}, while others simulate noisy quantum circuits or demonstrate error–mitigation techniques such as virtual distillation.  \change{Several notebooks cover importing and exporting measurement settings  and results to interface with quantum computing experiments using languages such as \texttt{qiskit} or \texttt{QASM}.}
Each notebook contains background material, references, and executable code illustrating data acquisition and post–processing workflows, serving as a practical complement to the examples presented in this article.  \change{Reproducible benchmark scripts are included in the repository under \texttt{benchmarks}.}

\section{Conclusion \& Outlook}\label{sec:outlook}

We have introduced \texttt{RandomMeas.jl}, a scalable and modular Julia package for developing, testing and implementing  randomized measurement (RM) protocols. Its architecture separates pre-processing, data acquisition and post-processing, enabling flexible integration with different experimental setups and simulation backends. The package supports the full RM workflow, from measurement setting generation to advanced estimators, and includes a tensor-network–based implementation of classical shadows suitable for large quantum systems. It offers robust routines for a variety of RM variants, built-in statistical error estimation, and a unified interface for both experimental and theoretical applications. These features make \texttt{RandomMeas.jl} well suited for current experiments and large-scale simulations, while providing a flexible foundation for future extensions.

Looking ahead, several development directions naturally emerge:

\paragraph{Machine learning and scalable tomography.}
Efficient and robust shadow-construction routines can be combined with classical machine learning techniques to design kernel methods for state classification~\cite{huang_provably_2022,lewis_improved_2024}, or to implement process shadow tomography for Liouvillian learning~\cite{kunjummen_shadow_2023,levy_classical_2024}. Scalable randomized measurements approaches to quantum state tomography, including methods based on MPOs~\cite{votto_learning_2026}, or on neural network quantum states~\cite{torlai_quantum_2023,huang_certifying_2024} (cf.\ the Julia package \href{https://github.com/GTorlai/PastaQ.jl}{PastaQ}), are another promising extension. Sampling strategies during pre-processing may be further optimized via importance sampling~\cite{hadfield_measurements_2020,rath_importance_2021} or derandomization techniques~\cite{huang_efficient_2021}, incorporating classical prior knowledge about the state or observable of interest \cite{elben_renyi_2018,Bringewatt_2024,Zhao_2024,sauvage2024classicalshadowssymmetries}.

\paragraph{Error mitigation.}
Randomized measurements enable several powerful quantum error mitigation techniques. \texttt{RandomMeas.jl} already supports robust shadows and can be naturally extended to include virtual distillation~\cite{seif_shadow_2023} and probabilistic error cancellation (PEC)~\cite{temme_error_2017,endo_practical_2018, takagi_shadow-pec_2023}. Such extensions could further leverage efficient tensor-network inversion techniques~\cite{filippov_scalable_2023} and symmetry restoration ideas based on symmetry-resolved shadows~\cite{vitale_symmetry-resolved_2022,joshi_observing_2024,Bringewatt_2024,Zhao_2024,sauvage2024classicalshadowssymmetries}, broadening applicability to realistic quantum devices.

\paragraph{Performance optimization with GPUs.}
While tensor-network methods in \texttt{RandomMeas.jl} already enable large-scale simulations, significant runtime gains could be achieved by leveraging GPU acceleration. In particular, GPU-based contraction of \texttt{ITensors.jl} objects would directly benefit the most computationally intensive post-processing routines.

By combining a standardized interface, modular design, and a focus on scalability and reproducibility, \texttt{RandomMeas.jl} provides a versatile platform for advancing the state of the art in randomized measurements. We invite contributions from the community to expand its functionality, integrate new protocols, and explore emerging applications at the interface of quantum information, simulation, and machine learning.

\section*{Acknowledgements}
We thank P.~Zoller, A.~Rath, V.~Vitale, R.~Kueng, H.~Y.~Huang and J.~Preskill for previous collaborations that triggered the development of this code. We thank L.~Nie, A.~Raj, A.~Vijay and J.~Kudler-Flam for discussions, in particular related to the estimation of trace moments.
BV thanks M.~Votto and D.~Lanier for feedback on the code.

\paragraph{Funding information}
Work in Grenoble is funded by the French National Research Agency via the research programs Plan France 2030 EPIQ (ANR-22-PETQ-0007), QUBITAF (ANR-22-PETQ-0004) and HQI (ANR-22-PNCQ-0002).

\bibliographystyle{quantum} 

\onecolumn\newpage
\appendix

\section{Jupyter notebooks for advanced usage}
\label{app:jupyter}

This appendix provides the complete list of Jupyter notebooks accompanying \texttt{RandomMeas.jl}.  
They can be found in the project repository, each with background material, references, and executable code.  
Many notebooks reproduce or reanalyse experimental results, while others are purely numerical and illustrate specific randomized–measurement protocols.

\subsection{Classical shadows}
\begin{enumerate}
    \item \textit{Energy measurements with classical shadows} —  
    Estimation of energies and energy variances, e.g.\ for verifying ground–state preparation in quantum simulation~\cite{kokail_self-verifying_2019, huang_predicting_2020}.
    \item \textit{Robust shadow tomography} —  
    Construction of classical shadows robust to measurement errors via calibration~\cite{chen_robust_2021,koh_classical_2022,vermersch_enhanced_2024,vitale_robust_2024}.
    \item \textit{Process shadow tomography} —  
    Illustration of quantum–gate fidelity estimation using common randomized measurements~\cite{kunjummen_shadow_2023,levy_classical_2024}.
    \item \textit{Classical shadows with shallow circuits} —  
    Calibration and use of shallow–circuit measurement settings for improved estimation of nonlocal observables~\cite{hu_classical_2023,akhtar_scalable_2023,bertoni_shallow_2024,hu_demonstration_2025}.
    \item \textit{Virtual distillation} —  
    Quantum error–mitigation technique based on classical shadows~\cite{seif_shadow_2023}.
\end{enumerate}

\subsection{Quantum benchmarking}
\begin{enumerate}
    \item \textit{Cross–entropy and self–cross–entropy benchmarking} —  
    Fidelity estimation for random circuits using classical simulation~\cite{arute_quantum_2019,andersen_thermalization_2025}.
    \item \textit{Fidelities from common randomized measurements} —  
    Estimation of $\braket{\psi|\rho|\psi}$ with reduced statistical errors~\cite{vermersch_enhanced_2024}.
    \item \textit{Cross–platform verification} —  
    Mixed–state fidelity estimation between states prepared on two quantum devices~\cite{elben_cross-platform_2020}.
\end{enumerate}

\subsection{Entanglement measurements}
\begin{enumerate}
    \item \textit{Entanglement entropy of pure states and the Page curve} —  
    Purity and Rényi–entropy estimation via randomized measurements.
    \item \textit{Analysis of Brydges et al.\ (Science 2019) data} —  
    Re–analysis of the first experimental randomized–measurement dataset~\cite{brydges_probing_2019,brydges_zenodo_2018}, including higher–order Rényi entropies and batch shadows~\cite{rath_entanglement_2023}.
    \item \textit{Mixed–state entanglement detection} —  
    Experimental detection of entanglement via the $p_3$–PPT condition~\cite{elben_mixed-state_2020}.
    \item \textit{Topological entanglement entropy} —  
    Simulation of purity measurements for the toric code and extraction of the topological contribution~\cite{satzinger_realizing_2021,elben_renyi_2018,brydges_probing_2019}.
\end{enumerate}

\subsection{Import/Export and interface with experiments}

\begin{enumerate}
    \item \change{ \textit{Qiskit hardware interface (example workflow)} — \texttt{RandomizedMeasurementsQiskit.ipynb} generates randomized measurement circuits and executes them in Qiskit. \texttt{RandomizedMeasurementsQiskitPostprocessing.ipynb} imports the exported unitaries and shots into \texttt{RandomMeas.jl} and performs the classical post-processing by constructing a \texttt{MeasurementGroup}.}
    
    \item \change{\textit{Exporting measurement circuits to OpenQASM} —  
Demonstration of the OpenQASM (QASM~2.0) export routine for \texttt{LocalUnitaryMeasurementSetting} and \texttt{ComputationalBasisMeasurementSetting}, enabling straightforward use of the corresponding randomized-measurement circuits in external SDKs (e.g.\ Qiskit, Cirq).}
\end{enumerate}

\subsection{Miscellaneous}
\begin{enumerate}
    \item \textit{Noisy circuit simulations with tensor networks} —  
    Simulation of noisy dynamics using density–matrix propagation or quantum–trajectory methods~\cite{jaschke_one-dimensional_2018}.
    \item \textit{Estimating statistical error bars via Jackknife resampling} —  
    Resampling–based uncertainty estimates from a single dataset.
\end{enumerate}

\end{document}